\documentclass[
times,
aps,
twocolumn,
showpacs,
superscriptaddress,
amsmath,
amssymb,
floatfix]{revtex4-2}

\usepackage{amsthm}
\usepackage{graphicx}
\usepackage{bm}
\usepackage[T1]{fontenc}
\usepackage{mathrsfs}
\usepackage{scrextend}
\usepackage{hyperref}
\usepackage{mathtools}
\usepackage{url}
\usepackage{braket}
\usepackage[dvipsnames]{xcolor}
\definecolor{C2}{RGB}{251, 77, 61}
\hypersetup{colorlinks=true, linkcolor=C2, citecolor=C2, urlcolor=C2}
\usepackage{dsfont}
\usepackage{epstopdf}

\newcommand{\e}{\mathrm{e}}
\renewcommand{\i}{\mathrm{i}}

\newcommand{\vac}{\mathrm{vac}}
\newcommand{\mbb}{\mathbb}

\usepackage{subcaption}
\usepackage{ragged2e}
\DeclareCaptionJustification{justified}{\justifying}

\begin{document}

\title{Quantum non-Markovianity elusive to interventions}

\author{Daniel Burgarth}
\email{daniel.burgarth@mq.edu.au}
\affiliation{Department of Physics and Astronomy, Center of Engineered Quantum Systems, Macquarie University, Sydney, NSW 2109, Australia}

\author{Paolo Facchi}
\email{paolo.facchi@ba.infn.it}
\affiliation{Dipartimento di Fisica and MECENAS, Universit\`{a} di Bari, I-70126 Bari, Italy}
\affiliation{INFN, Sezione di Bari, I-70126 Bari, Italy}

\author{Davide Lonigro}
 \email{davide.lonigro@ba.infn.it}
\affiliation{Dipartimento di Fisica and MECENAS, Universit\`{a} di Bari, I-70126 Bari, Italy}
\affiliation{INFN, Sezione di Bari, I-70126 Bari, Italy}

\author{Kavan Modi}
\email{kavan.modi@monash.edu}
\affiliation{School of Physics and Astronomy, Monash University, Clayton, Victoria 3800, Australia}

\date{\today}

\begin{abstract}
The non-Markovian nature of open quantum dynamics lies in the structure of the multitime correlations, which are accessible by means of interventions. Here, by examining multitime correlations, we show that it is possible to engineer non-Markovian systems with only long-term memory but seemingly no short-term memory, so that their non-Markovianity is completely non-detectable by any interventions up to an arbitrarily large time. Our results raise the question about the assessibility of non-Markovianity: in principle, non-Markovian effects that are perfectly elusive to interventions may emerge at much later times.
\end{abstract}

\maketitle

\textbf{Introduction.---} A key obstacle that lies in the path to quantum information processing is noise~\cite{13preskill}. The conventional models for quantum noise, responsible for decoherence of qubits, make many simplifying assumptions. One of the key assumption is that the noise is memoryless or Markovian~\cite{helsen2020general}; this is widely known to be false, and an immense effort in understanding non-Markovian noise, both in general and in quantum information processors, has been initiated~\cite{Rivas2014, Li2018, modi}. While non-Markovian noise is more complex than Markovian, it is \emph{not} more detrimental. In fact, non-Markovian effects, which manifest as correlations in time, can be used to improve the functionality of the quantum information processor~\cite{PhysRevLett.82.2417, Arenz2017, White2020}. Thus, modelling and characterising different variety of non-Markovian noise is of strong interest.

The first challenge in this endeavour is to be able to differentiate between Markovian and non-Markovian noise in the quantum regime, which is not an easy task. Often, Markovian noise is associated with the exponential decay curves, e.g. a qubit that relaxes to the maximally mixed state exponentially fast. However, there are instances where a qubit exhibits an exponential decay, but nevertheless is undergoing a non-Markovian process~\cite{Lindblad1980, accardi}. A famous example is due to Lindblad, dubbed as \textit{shallow pocket} (SP), which has been scrutinised in detail recently in terms of dynamical decoupling~\cite{PhysRevA.92.022102,Milz2019}, signalling~\cite{Milz2019}, and multitime correlations~\cite{Phil_MemStr}. (See Ref.~\cite{arXiv:2010.00279} for a generalisation of SP.) In each, case it is clear that the seemingly simple Markovian noise is, in fact, complex non-Markovian noise that can be exploited to prolong the coherence time of the system. On the other hand, there is a class of system-environment dynamics, generated by Chebotarev-Gregoratti (CG) Hamiltonians~\cite{10.1007/s002200100500} (also see Ref.~\cite{arXiv:1904.03627}), that yield an exponential decay for the system but, unlike SP, do not allow for dynamical decoupling or other noise mitigation methods.

Recently, the matter took a turn for worse; Ref.~\cite{HidNM} introduced a class of models in which the decay of the system exhibits the tantalisingly Markovian exponential decay for a finite, but arbitrarily large, time window; then, abruptly, there is a departure from the exponential decay, i.e., the system's evolution displays non-Markovian features of the noise. These models have thus been termed as \textit{hidden non-Markovian} (HNM) models. They occupy the space somewhere between SP and CG models. Importantly, given one of these process, a snapshot~\cite{PhysRevLett.101.150402} of the dynamics can fail to differentiate between them and will label each as Markovian~\cite{Rivas2014,Li2018}. This is an important space, as most real experiments exhibit the aforementioned exponential decay, and suggests that non-Markovianity may be unassessibile.

One possibility is to examine how the system reacts to interventions. Indeed, since non-Markovian systems tend to respond positively to error mitigation methods such as dynamical decoupling, it would appear natural that even models with hidden non-Markovianity can be `stimulated' by interventions to already reveal this at much earlier time. For instance, with the right type of intervention we might be able to fill up the quantum environment with excitation and stimulate an earlier backflow. Multitime correlation measurements were also shown to reveal the failure of the quantum regression theorem, and therefore a subtle type of non-Markovianity~\cite{Guarnieri2014}, which may appear as a precursor.

In this Letter we show that the class of HNM processes are, in fact, genuinely Markovian within the finite window and genuinely non-Markovian outside of that temporal window. We do this by computing the multitime correlations, including those stemming from interventions. This is provably a set of \textit{necessary and sufficient} quantum Markov conditions~\cite{PhysRevA.97.012127, PhysRevLett.120.040405, Costa_2016}. We begin by revisiting the role of multitime correlations in quantum stochastic processes, and the basic properties of HNM models; then we compute such correlations for HNM models, showing that they vanish identically at small times. In other words, we show that there are non-Markovian processes whose non-Markovianity is \emph{perfectly hidden} for a finite duration; non-Markovianity can be therefore undetectable at small times, even by taking into account the role of interventions, since the onset of all time correlations can be arbitrarily delayed.\smallskip

\textbf{Quantum multitime correlations.---} A classical stochastic process is a joint probability distribution of a random variable $x$ in time. In practice, one usually considers the joint distribution of a discrete set of times, $\{t_k, \dots , t_0\}$ and corresponding probability distributions of $\mbb{P}(x_k,\dots,x_0)$. A Markovian process satisfies the following condition: $\mbb{P}(x_k|x_{k-1}, \dots,x_0) =\mbb{P}(x_k|x_{k-1})$, i.e., the system, at a given time, is only conditionally dependent of its state at the previous time step. Practically speaking, this condition means that a Markov process is easy to work with. Nevertheless, such processes are not the norm, but rather a special case~\cite{vanKampen1998}.

\begin{figure*}[t!]
\begin{subfigure}[]{0.69\linewidth}
\includegraphics[width=.95\linewidth]{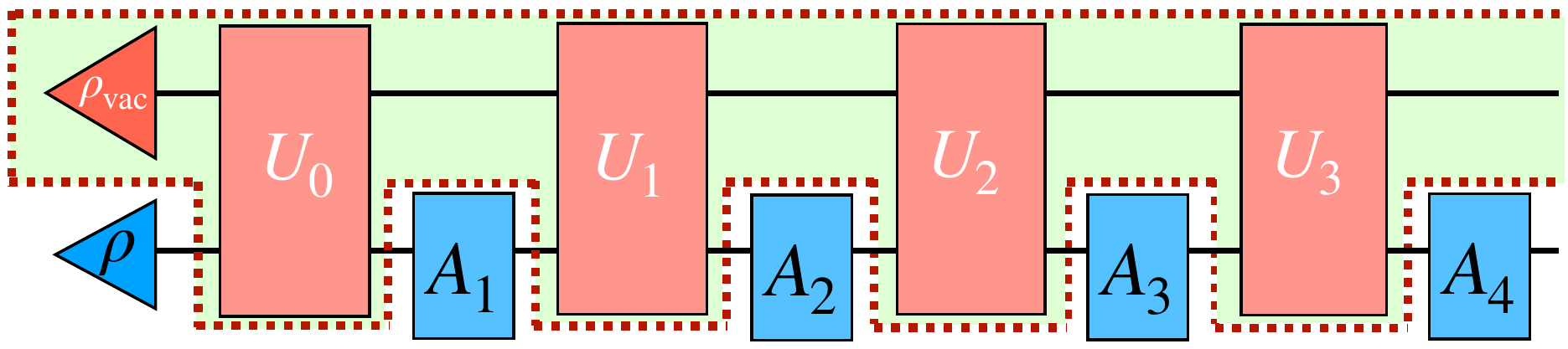}
\caption{}\label{fig:b}\vspace{0.43cm}
\includegraphics[width=.95\linewidth]{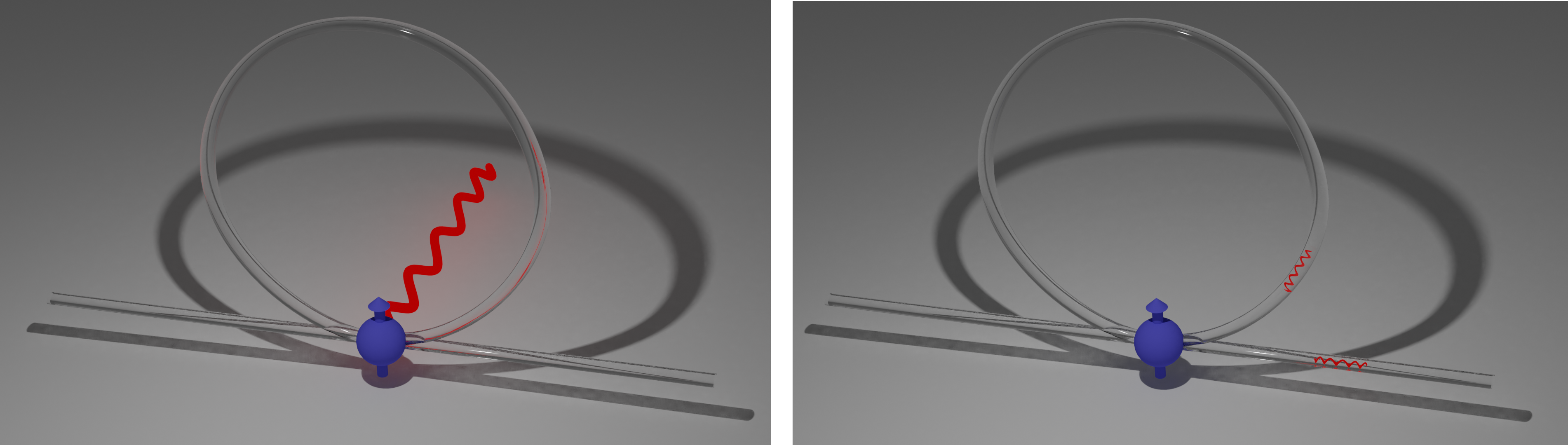}
\caption{}
\label{fig:a}
\end{subfigure}\hfill
\begin{subfigure}[]{0.29\linewidth}
\includegraphics[width=\linewidth]{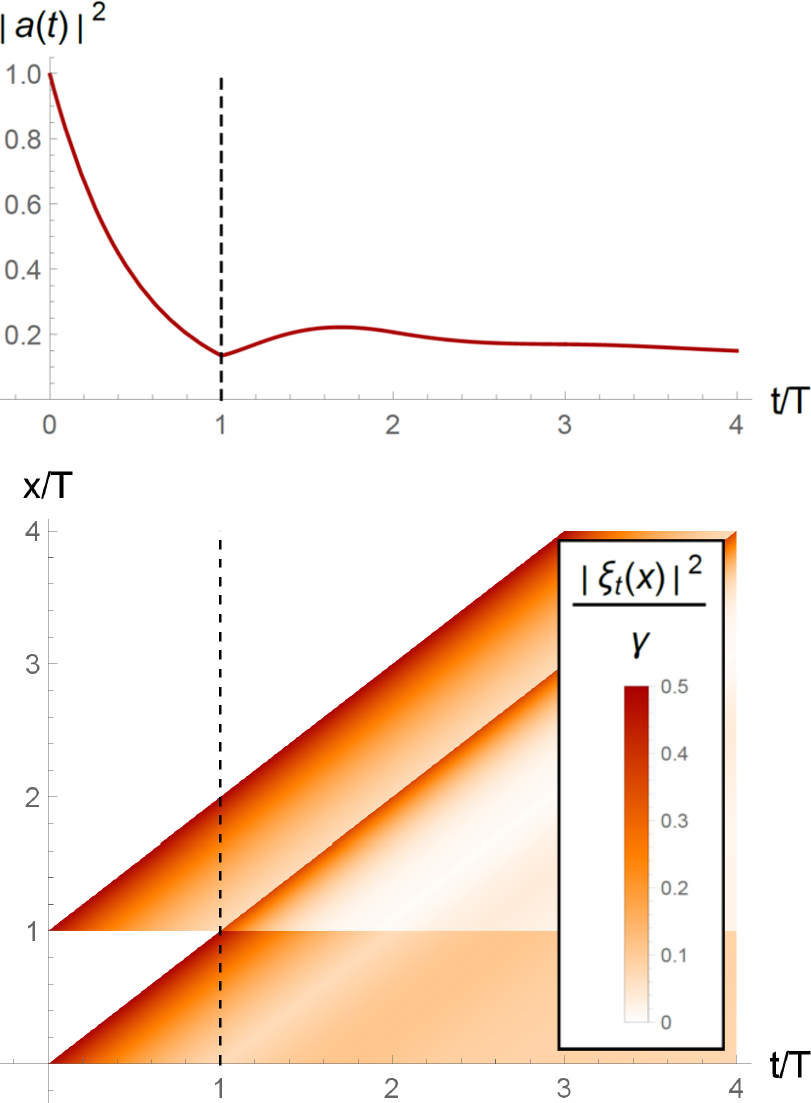}
\caption{}\label{fig:c}
\end{subfigure}
\captionsetup{justification=justified, singlelinecheck=false}
\caption{(a): Scheme of a quantum stochastic process for a system coupled with an environment, initially prepared in a global state $\varrho_0=\rho\otimes\rho_{\rm vac}$. At times $t_1,\dots,t_4$, an intervention is performed on the system alone; between consecutive interventions, system and environment evolve freely. (b): Scheme of a qubit in a photon waveguide with a two-point interaction. The loop of length $T$ creates a time-delayed coherent feedback and non-Markovian effects on the qubit evolution. (c): Free evolution of $\Ket{0,\mathrm{vac}}$: plot of the survival probability of the emitter $|a(t)|^2$ (top) and density plot of the wavefunction of the emitted boson $|\xi_t(x)|^2$ (bottom) for the model with double point interaction, with $\gamma T=\omega_0T=2$; darker shades correspond to larger values of $|\xi_t(x)|^2$. In both graphs, the dashed black line represents the time $t=T$ at which $|a(t)|^2$ ceases to be exponential and Eq.~\eqref{eq:singlephoton2} ceases to hold.}
\label{allfigs}
\end{figure*}

For multitime correlations in the quantum case, consider an initial system-bath ($SB$) state $\varrho_0$, which undergoes an evolution $U_0$ and then an intervention $A_0$ is then made on the system $S$ alone. Next, the total state once again evolves due to $U_1$, followed by a second intervention $A_1$ on $S$ alone, and so on up to a final intervention $A_k$ is performed following $U_k$, see Fig.~\ref{fig:b}. The interventions $\{A_j\}$ are any physically implementable operation, which can be thought of a generalised measurement with possible corresponding outcomes $\{x_j\}$. Mathematically, these are known as \textit{instruments}~\cite{davies} and represented by a collection of completely positive  map $\mathcal{J} := \{A_{x_j}\}$ such that $\sum_{x_j} A_{x_j}$ is  trace preserving.

The above machinery straightforwardly allows for the calculation of the probability to observe a sequence of quantum events $(x_k, \dots x_0)$, corresponding to a choice of instruments $\{\mathcal{J}_k, \dots, \mathcal{J}_0\}$, as
\begin{gather}\nonumber
\mbb{P}(x_k,\dots,x_0\,|\,\mathcal{J}_k,\dots, \mathcal{J}_0)= \mbox{tr}[{A}_{x_k} {U}_{k-1} \!\cdots {A}_{x_0} {U}_0 (\varrho_0)].
\end{gather}
While the LHS is akin to a classical joint probability distribution, we have yet to identify the quantum stochastic process. We can do this by rewriting the RHS as
\begin{gather}
\mbox{tr}[{A}_{x_k} {U}_{k-1} \!\cdots {A}_{x_0} {U}_0 \varrho_0] = \mbox{tr}[\Upsilon_{k:0} \mathbf{A}^{\rm T}_{k:0}] \label{eq:process}
\end{gather}
with $\mathbf{A}_{k:0} \!:=\! {A}_{x_k} \!\otimes\! \dots \otimes {A}_{x_0}$ and $\Upsilon_{k:0} \!:=\! \mbox{tr}_B [{U}_k \star \dots \star {U}_0 \varrho_0]$, where ${\rm T}$ denotes transposition and $\star$ denotes the link product, defined as a matrix product on the space $B$ and a tensor product on space $S$~\cite{PhysRevA.80.022339}. The important feature here is the clear separation of the interventions $\mathbf{A}_{k:0}$ from the influences due to the bath, which are packaged in $\Upsilon_{k:0}$, which is the \emph{Choi state} of the process~\cite{modi}.

The process tensor $\Upsilon_{k:0}$ is the quantum generalisation of the joint classical probability distribution and unambiguously represents a quantum stochastic process~\cite{Quolmogorov, PhysRevA.100.022120, arXiv:1907.05807}. It contains all accessible multitime correlations~\cite{White2020, White2021a}, including temporal entanglement~\cite{arXiv:1811.03722, 10.21468/SciPostPhys.10.6.141, White2021b}, and Markovian processes are those satisfying the following property: any $k$-time process tensor factorises as
\begin{gather} \label{eq:markovchoi}
\Upsilon_{k:0} = \Upsilon_{k:k-1} \otimes \cdots \otimes \Upsilon_{2:1} \otimes  \Upsilon_{1:0}.
\end{gather}
Conversely, we can deduce a process to be non-Markovian by looking for correlations. For instance, in the SP model, while the qubit dephases exponentially, an intervention of a Pauli $\sigma_x$ operation reverses this process: therefore, in this model non-Markovianity isn't seen in two-time correlations, but it is seen in three-time correlations.\smallskip

\textbf{Hidden non-Markovianity models.---} We shall consider a qubit with ground state $\ket{1}$ and excited state $\ket{0}$, with excitation energy $\omega_0$, interacting with an one-dimensional bosonic bath on the real line. We write the Hamiltonian of the bath in the position representation, with   annihilation and creation operators $b_x$, $b_x^\dag$ satisfying the commutation relations $[b_x,b_{x'}^\dag]=\delta(x-x')$. Throughout this section, we shall use the following compact notation: for any wavefunction $\eta(x)$, we set
\begin{equation}\label{eq:op}
    B^\dag(\eta)=\int\mathrm{d}x\;\eta(x)\, b_x^\dag,
\end{equation}
which represents the creation operator of a boson with wavefunction $\eta(x)$. The total Hamiltonian $H= H_0 + H_{\mathrm{int}}$ is the sum of the free Hamiltonian
\begin{equation}\label{eq:h0}
	H_0 = \omega_0\ket{0}\!\bra{0} \otimes \mathbf{1} + \mathbf{1}\otimes H_B,
\end{equation}
and the interaction Hamiltonian
\begin{equation}\label{eq:h1}
	H_{\mathrm{int}} = \sigma_+ \otimes B(g)+\sigma_- \otimes B^\dag(g).
\end{equation}
Above, $H_B$ is the second quantisation of the momentum operator $p=-\i\frac{\mathrm{d}}{\mathrm{d}x}$ on the boson field, $\sigma_+=\sigma_-^\dag=\ket{0}\!\bra{1}$, and $g(x)$ is the form factor which encodes all information about the qubit-field interaction; different choices of the form factor can yield drastically distinct physics.

\emph{Local point interaction.---}
A point interaction between the field and a qubit at $x=0$ is obtained by setting $g(x)=\sqrt{\gamma}\,\delta(x)$, with coupling constant $\gamma>0$, which gives
$B(g)=\sqrt{\gamma} \, b_0$ \footnote{Strictly speaking, the sum of~\eqref{eq:h0} and~\eqref{eq:h1} is a well-defined (and self-adjoint) Hamiltonian only when $g(x)$ is  square integrable, which is clearly not the case here; however, as discussed in~\cite{fl1,fl2}, via a renormalisation procedure (\emph{singular coupling limit}) a self-adjoint operator can be obtained even in the cases that we are considering.}.
This choice corresponds to a flat form factor in the frequency representation used in Ref.~\cite{HidNM}. In the single-excitation sector, the model is exactly solvable; in particular, as shown in the Appendix, for every $t\in\mathbb{R}$ we have $e^{-\i tH}\ket{1,\vac} =\e^{-\i \varepsilon_0t}\ket{1,\vac}$ and
\begin{equation}
	e^{-\i tH}\ket{0,\vac}=a(t)\ket{0,\vac}+B^\dag(\xi_t)\ket{1,\vac}.
	\label{eq:spontdecay}
\end{equation}
Above, $\ket{s,\vac}\equiv\ket{s}\otimes\ket{\vac}$, $s=0,1$, with $\ket{\vac}$ being the field vacuum state; besides, $a(t)=e^{-\left(\i \varepsilon_0+\frac{\gamma}{2}\right)t}$ with $\varepsilon_0$ being the dressed excitation energy of the qubit; finally, the boson wavefunction is given by $\xi_t(x)= \varphi_t(x)$, where 
\begin{equation}\label{eq:singlephoton}
\varphi_t(x)=-\i\sqrt{\gamma}\, a(t-x)\, 1_{[0,t]}(x),
\end{equation}
with $1_{\Omega}(x)$ being the characteristic function of the set $\Omega$, that is, $1_{\Omega}(x)=1$ for $x\in\Omega$ and $=0$ elsewhere. This is a simple model of an emission process: the decay of the qubit at $x=0$ is associated with the creation of a photon which propagates along the positive direction of the $x$ axis with unit velocity. The photon wavefunction exactly traces out the exponential decay of the qubit and is compactly supported in the interval $[0,t]$.

Consequently, the quantum evolution $\Lambda_t$ of the qubit obtained by preparing the system in a state $\rho$ and the field in the vacuum $\rho_{\vac}=\Ket{\mathrm{vac}}\!\Bra{\mathrm{vac}}$, letting them evolve  for a time $t$, and tracing out the field, namely
\begin{equation}\label{eq:process1}
\rho(t)\equiv\Lambda_t(\rho)=\mbox{tr}_B \bigl[\e^{-\i tH}\rho\otimes\rho_\vac\,\e^{\i  tH}\bigr],
\end{equation}
satisfies the semigroup property $\Lambda_t \Lambda_s = \Lambda_{t+s}$ at all times, and yields a Markovian evolution: indeed, it is simply an amplitude-damping channel with decay rate $\gamma$~\cite{HidNM}.

\emph{Nonlocal point interactions.---}
Now we consider a modification of the above setup. Namely, we  allow for a nonlocal interaction of the qubit at two distinct points, say $x=0$ and $x=T$, given by a form factor 
\begin{equation}\label{eq:formfactor}
	g(x)=\sqrt{\frac{\gamma}{2}}\,\bigl(\,\delta(x) + \delta(x-T)\,\bigr),
\end{equation}
so that $B(g)=\sqrt{\gamma/2} \, (b_0 + b_T)$. See Fig.~\ref{fig:a}. Physically, such a choice (up to a relative sign) has been used, see e.g.~\cite{Fang_2018}, as a  model of a qubit in a single-end waveguide with a perfect mirror at one end, with $T/2$ being the distance between the qubit and the mirror. This model was also studied in Ref.~\cite{HidNM} as an example of non-Markovian model yielding exponential decay for the reduced dynamics of the qubit up to a time $T$.

Again, the single-excitation sector is fully solvable (see Appendix). In particular, for $0\leq t\leq T$, we have again an evolved state~\eqref{eq:spontdecay} with $a(t)=\exp\left(-\left(\i \varepsilon_0+\frac{\gamma}{2}\right)t\right)$, and 
 \begin{equation}\label{eq:singlephoton2} 	\xi_t(x)=\frac{1}{\sqrt{2}}\bigl(\,\varphi_t(x) + \varphi_{t}(x-T)\,\bigr),\quad t\in[0,T],
 \end{equation}
where $\varphi_t(x)$ is as in Eq.~\eqref{eq:singlephoton}. We may interpret this new situation as a two-point emission: at time $t=0$, the qubit emits a photon at both positions $x=0,T$, and each of its two spatially separated part propagates separately in the positive direction of the $x$ axis up to the time $t=T$. See Fig.~\ref{fig:a}. Notice that the two components, up to normalisation, are two exact copies of the wavefunction of the one-point emission, and in particular the overall norm $\|\xi_t\|^2=\|\varphi_t\|^2=1-e^{-\gamma t}$ is the same as before. In this sense, at times smaller than $T$, this system behaves like the superposition of two identical, independent ``copies'' of the previous system. When $t=T$, one of the photon branches starts interfering back with the qubit and this simple picture ceases immediately to hold: $a(t)$ is no longer an exponential function, and $\xi_t(x)$ will no longer satisfy Eq.~\eqref{eq:singlephoton2}. This is clearly visible in the plots of $|a(t)|^2$ and $|\xi_t(x)|^2$ in Fig.~\ref{fig:c}.

As a result, the evolution of the qubit state $\rho(t)$, obtained in the same way as in Eq.~\eqref{eq:process1} will be completely indistinguishable from the previous one, as long as $0\leq t\leq T$. In fact, the form factor given by Eq.~\eqref{eq:formfactor} is just one possible example: the same result would be obtained by taking into account any number of spatially separated nonlocal point interactions: until the various components of the photon wavefunction do not propagate to the next point and start interfering. We can thus construct a large family  of quantum systems with the HNM property, as shown in Ref.~\cite{HidNM}. In all such cases, we obtain non-Markovian processes which, as long as we examine their free dynamics, ``look exactly like'' a Markovian process at times $0\leq t\leq T$, despite ultimately starting to show non-Markovian behaviour at $t>T$.

This argument, however, still does not take into account multitime correlations of the qubit, which are accessible by means of external interventions on the system. The above result only accounts for correlations between two times, and in general, correlations at \textit{all} orders should be proved to  vanish for assessing that a process is Markovian.

We will make a few remarks before going on and examining multitime correlations. The spin-boson model~\eqref{eq:h0}--\eqref{eq:h1}, with a singular point interaction $g(x)=\sqrt{\gamma}\,\delta(x)$ and with boson Hamiltonian $H_B$ the right shift generator, can be mapped into an SLH model~\cite{10.1080/23746149.2017.1343097, 10.1007/s002200100500}, with $S=1$, $L=\sigma_-$, and $H=\varepsilon_0 \ket{0}\bra{0}$. As such, it is associated with an It\^{o} quantum stochastic differential equation~\cite{10.1007/BF01258530}, with a regression property for all multitime correlation functions, and thus with a strong quantum Markovianity property~\cite{Li2018}. In this respect, the model with a two-point interaction~\eqref{eq:formfactor}, is a non-Markovian generalisation of an SLH model, in which a feedback is considered, see Figure~\ref{fig:a}. This is also linked with the theory of time-delayed coherent quantum feedback~\cite{PhysRevLett.115.060402}, where a similar scheme was considered. Interestingly enough, this theoretical model is an effective description of the experimental implementation of superconducting artificial giant atoms in waveguide electrodynamics~\cite{10.1038/s41586-020-2529-9}.\smallskip

\textbf{Multitime correlations in HNM Models.---} Now we turn to the main question of this Article, that is, whether higher order correlations are present in HNM models. In other words, can intermediate interventions reveal hidden non-Markovianity already in the time interval $[0,T]$? Such interventions can cause the system to go in higher-excitation sectors, which are not solvable; the ``indistiguishability'' between this model and the reference one may, in principle, be broken. 

However, as shown in the Appendix, the system with form factor~\eqref{eq:formfactor}  satisfies a property which will be crucially employed here: given a time $t<T$ and any wavefunction $\eta(x)$ which satisfies
\begin{equation}\label{eq:supp}
    \eta(x) =0 , \quad \text{for } x\in [-t,0]\cup[T-t,T],
\end{equation}
then the following property holds for $B^\dag(\eta)$ in~\eqref{eq:op}:
\begin{equation}\label{eq:commute2}
    \e^{-\i tH}B^\dag\left(\eta\right)\e^{\i tH}=B^\dag\!\left(\e^{-\i tp}\eta\right),
\end{equation}
where $\e^{-\i tp}\eta(x)=\eta(x-t)$ is  the free evolution of $\eta(x)$. Physically, Eq.~\eqref{eq:commute2} can be interpreted as follows: since wavefunctions satisfying~\eqref{eq:supp} are sufficiently far apart from the point interactions, they freely propagate without interfering.
In particular, this implies that, for $t_1,t_2\geq0$ such that $t_1+t_2<T$, we have
\begin{equation}\label{eq:commute3}
    \e^{-\i t_2H}B^\dag(\xi_{t_1}) \e^{\i t_2H}=B^\dag\!\left(\e^{-\i t_2p}\xi_{t_1}\right),
\end{equation}
with $\xi_t(x)$ being the boson wavefunction~\eqref{eq:singlephoton2}. This simple equation encodes a fundamental property of the system, which proves to be the key point of our discussion: as long as the total time of observation is less than $T$, the boson field \emph{cannot carry information} about the non-Markovianity of the system. Local operations of any kind on the qubit cannot modify this simple picture: however we intervene on the qubit via a process as depicted in Fig.~\ref{fig:b}, with all interventions made at times $t_j<T$, all photons emitted in the process will evolve exactly as they would in the absence of coupling. Consequently, all multitime correlations vanish, and the process is genuinely Markovian up to a time $T$. Only at later times the field will ``recognise'' the inherently non-Markovian structure of the coupling. 

This  argument is fully backed up by a lengthy but straightforward calculation which shows that, as a consequence of property~\eqref{eq:commute3}, the process tensor $\Upsilon_{k,0}$ for all the HNM models (and, in particular, for the two models discussed in the previous section) is \emph{exactly the same}, and has the factor form~\eqref{eq:markovchoi}, as long as $t_k\in[0,T]$. However, for $t_1\in[0,T]$ and $t_2 >T$, the process tensor will display correlations. Therefore, all HNM models  define a quantum process which is genuinely Markovian within a finite time window and (apart from the one-point interaction which is Markovian for all times) genuinely non-Markovian outside that window. The proof is shown in the Appendix for three-time correlations, and may be easily generalised for an arbitrary number of interventions.

What is surprising here is that these intricate memory structures stem from a rather simple time-independent Hamiltonian. This is a feature that may be used to engineer intricate temporal correlations, which we discuss in our concluding remarks.\smallskip

\textbf{Conclusions.---} When the Markovian properties of quantum noise were first investigated~\cite{Lindblad1980,accardi} the emphasis was on the natural properties of a given system. Nowadays, quantum information adds an important engineering perspective, which asks how systems behave differently in the context of design and control. This means that device characterisation needs to test quantum systems under a wide range of interventions. We have shown here that it is ultimately impossible, in general, to fully conclude if a given system is truly Markovian, no matter how complex this characterisation is. Using such a device under Markovian design assumptions can then lead to unexpected behaviour at a later stage.

On the other hand, open systems quickly become too complex. Simple models like SP, CG, and now HNM, allow us to form simple building block for complex processes and offer keen intuition about the structures of quantum stochastic processes. There are several features of the HNM of interest. Firstly, this is an example of a process which has no short-term memory and only long-term memory. Moreover, this model could serve as the basis for constructing processes with only higher-order correlations, that is, a process where only correlations above four points in time are nonvanishing. Finally, HNM can serve as an Ansatz for simulating processes with slow decaying correlations. To do this, we may add more loops to at the top in Fig.~\ref{fig:a} and create a self-similar structure that will reduce the correlation strength geometrically after a delay of $T$. The slowly decaying correlation in time here will be akin to the highly common $1/f$ noise. Modelling processes with slowly decaying noise is thought to be hard for the same reasons as the tensor network representation of such correlation is non-trivial~\cite{orus_practical_2014}.

\begin{acknowledgments}
DB acknowledges funding by the ARC (project numbers FT190100106, DP210101367, CE170100009). PF and DL were partially supported by the Italian National Group of Mathematical Physics (GNFM-INdAM), by Istituto Nazionale di Fisica Nucleare (INFN) through  the  project  “QUANTUM”, and by Regione Puglia and QuantERA ERA-NET Co-fund in Quantum Technologies (GA No. 731473), project PACE-IN. KM is supported through Australian Research Council Future Fellowship FT160100073, Discovery Project DP210100597, and the International Quantum U Tech Accelerator award by the US Air Force Research Laboratory.
\end{acknowledgments}

\appendix

\section{}

\subsection{Dynamics in the single-excitation sector}\label{app:singleexcitation}
As discussed in the main text, the Hamiltonian given by $H=H_0+H_{\mathrm{int}}$, with $H_0$ as in Eq.~\eqref{eq:h0} and $H_{\mathrm{int}}$ as in Eq.~\eqref{eq:h1}, preserves the total number of excitations of states. Let us consider the single-excitation sector, spanned by states in the form
\begin{equation}\label{eq:singleexc}
	\ket{\Psi}=a\ket{0,\vac}+B^\dag(\xi)\ket{0,\vac}
\end{equation}
for some $a\in\mathbb{C}$ and some wavefunction $\xi(x)$, with $B^\dag(\xi)$ being the operator that creates a photon with wavefunction $\xi$, as defined in Eq.~\eqref{eq:op}. 

The restriction of $H$ to vectors in the form~\eqref{eq:singleexc}, known as the Friedrichs-Lee Hamiltonian, is exactly solvable; see Refs.~\cite{fl1,fl2}. We will be interested in the evolution of the states $\ket{1,\vac}$ and $\ket{0,\vac}$; the first one is readily shown to evolve trivially, for the second one we need to solve the Schr\"odinger equation, where we set $\xi_t(x)=\xi(t,x)$:
\begin{equation}\label{eq:schro}
	\begin{dcases}
	\i\,\dot a(t)=\omega_0a(t)+\int\mathrm{d}x\,\overline{g(x)}\xi(t,x);\\
	\i\,\dot \xi(t,x)=-\i\,\xi'(t,x)+a(t)g(x)
	\end{dcases}
\end{equation}
with initial conditions $\xi(0,x)\equiv0$ and $a(0)=1$, with $\dot\xi(t,x)$, $\xi'(t,x)$ being respectively the derivatives with respect to $t$ and $x$. 

While this system has already been solved in Ref.~\cite{HidNM}, we will briefly discuss here its resolution since we are now working in the (different, but equivalent) position representation. The second equation with initial condition at $t=0$ can be solved formally with respect to the first one:
\begin{equation}\label{eq:formal}
	\xi(t,x)=-\i \int_0^t\mathrm{d}s\;a(s)g(x-(t-s)).
\end{equation}
In the main text we have considered the following two choices:
\begin{itemize}
	\item $g(x)=\sqrt{\gamma}\,\delta(x)$;
	\item $g(x)=\sqrt{\frac{\gamma}{2}}\left(\delta(x) + \delta(x-T)\right)$.
\end{itemize}
In the first case, Eq.~\eqref{eq:formal} simplifies as follows:
\begin{eqnarray}\label{eq:photon1}
	\xi(t,x)&=&-\i \sqrt{\gamma}\int_0^t\mathrm{d}s\;a(s)\delta(x-(t-s))\nonumber\\
	&=&-\i \sqrt{\gamma}\,a(t-x)\,1_{[0,t]}(x),
\end{eqnarray}
where $1_{[0,t]}(x)$ is the characteristic function of the interval $[0,t]$, i.e. $1_{[0,t]}(x)=1$ for $x\in[0,t]$ and is zero otherwise. Substituting this expression in the first equation in~\eqref{eq:schro}, it can be shown, as discussed extensively in the Appendix of Ref.~\cite{HidNM}, that $a(t)$ is a pure exponential function:
\begin{equation}\label{eq:exp}
a(t)=e^{-\left(\i \varepsilon_0+\frac{\gamma}{2}\right)t}
\end{equation}
with $\varepsilon_0$ being the dressed (renormalised) excitation energy of the qubit. See Refs.~\cite{fl1,fl2} for a rigorous approach to the renormalisation in the Friedrichs-Lee model. By substituting Eq.~\eqref{eq:exp} into Eq.~\eqref{eq:photon1}, one obtains
\begin{equation}
	\xi_t(x)\equiv\varphi_t(x)=-\i \sqrt{\gamma}e^{-\left(\i \varepsilon_0+\frac{\gamma}{2}\right)(t-x)}1_{[0,t]}(x),
\end{equation}
as reported in Eq.~\eqref{eq:singlephoton} in the main text. Notice that
\begin{equation}
	\|\varphi_t\|^2=1-e^{-\gamma t}.
\end{equation}
In the second case, Eq.~\eqref{eq:formal} becomes
\begin{widetext}
\begin{eqnarray}\label{eq:photon2}
\xi_t(x)&=&-\i \sqrt{\frac{\gamma}{2}}\int_0^t\mathrm{d}s\;a(s)\left(\delta(x-(t-s))+\delta(x-(t-s+T))\right)\nonumber\\
&=&-\i \sqrt{\frac{\gamma}{2}}\left(a(t-x)\,1_{[0,t]}(x)+a(t+T-x)1_{[T,T+t]}(x)\right)\nonumber\\
&=&-\i \sqrt{\frac{\gamma}{2}}\left(a(t-x)\,1_{[0,t]}(x)+a(t-(x-T))1_{[0,t]}(x-T)\right).
\end{eqnarray}
\end{widetext}
Notice that $\xi_t(x)$ is the sum of two identical, compactly supported wavefunctions evaluated at $x$ and $x-T$ respectively; in particular, for $t<T$ their supports are disjoint. See also Fig.~\ref{fig:c} in the main text.
 
Substituting this expression in the first equation in~\eqref{eq:schro}, again $a(t)$ can be computed exactly at all times, as discussed in Ref.~\cite{HidNM}, while its expression will be involved for $t>T$, for $t\leq T$ it is, again, a pure exponential given by Eq.~\eqref{eq:exp}, and therefore
\begin{equation}
\label{eq:boson}
	\xi_t(x)=\frac{1}{\sqrt{2}}\left(\varphi_t(x)+\varphi_{t}(x-T)\right),
\end{equation}
again with $\varphi_t(x)$ given by Eq.~\eqref{eq:singlephoton}. Notice that, for $t<T$,
\begin{eqnarray}\label{eq:bosonnorm}
	\|\xi_t\|^2&=&\frac{1}{2}\left(\|\varphi_t\|^2+\|\varphi_{t+T}\|^2\right)\nonumber\\
	&=&\|\varphi_t\|^2=1-e^{-\gamma t},
\end{eqnarray}
again as an immediate consequence of the fact that, for $t<T$, the wavefunction is composed of two identical, disjointly supported terms.

Finally, we remark that the generalisation of these calculations to the case of a form factor 
\begin{equation}\label{eq:generic}
	g(x)=\sum_nc_n\delta(x-x_n),\quad\sum_n|c_n|^2=1, 
	\quad x_{n+1}-x_n \geq T,
\end{equation}
is straightforward: for each admissible choice of the coefficients $c_n$ and the points $x_n$, the function $a(t)$ will again be a pure exponential up to $T$ and the resulting single-photon wavefunction $\xi_t(x)$ will be a superposition of rescaled and translated components $c_n \varphi_t(x-x_n)$; again, the property $\|\xi_t\|^2=\|\varphi_t\|^2$ will hold.

\subsection{Proof of Eq.~\eqref{eq:commute2}}\label{app:commute}
We will now show that Eq.~\eqref{eq:commute2} holds, respectively in the following cases:
\begin{itemize}
	\item if $g(x)=\sqrt{\gamma}\,\delta(x)$, whenever $\eta$ is not supported in $[-t,0]$;
	\item if $g(x)=\sqrt{\frac{\gamma}{2}}\bigl(\,\delta(x)+\delta(x-T)\,\bigr)$, whenever $\eta$ is not supported in $[-t,0]\cup[T-t,T]$,
\end{itemize}
the generalisation to a form factor~\eqref{eq:generic} again being immediate. 

Let $B^\dag(\eta,t)=\e^{-\i tH}B^\dag(\eta)\e^{\i tH}$ the Heisenberg evolution of $B^\dag(\eta)$; as such, it must satisfy
\begin{equation}
\begin{cases}
    \frac{\mathrm{d}}{\mathrm{d}s} B^\dag(\eta,s)=-\i [H,B^\dag(\eta,s)] ,\\
	B^\dag(\eta,0)=B^\dag(\eta).
\end{cases}	
\end{equation}
On the other hand, the free evolution $B^\dag(\e^{-\i s p}\eta)$ satisfies
\begin{equation}
\begin{cases}
    \frac{\mathrm{d}}{\mathrm{d}s} B^\dag(\e^{-\i s p}\eta)=-\i [H_B, B^\dag(\e^{-\i s p}\eta)] , \\
	B^\dag(\e^{-\i s p}\eta)\Big |_{s=0}=B^\dag(\eta),
\end{cases}	
\end{equation}
where the free boson Hamiltonian $H_B$ is the second quantisation of $p$.
Now, in the case of a point interaction at $x=0$, we get
\begin{align}
    [H,B^\dag(\e^{-\i s p}\eta)] &= [H_B,B^\dag(\e^{-\i s p}\eta)] + \sqrt{\gamma}\, (\e^{-\i s p}\eta)(0) \sigma_+
    \nonumber\\
    &= [H_B,B^\dag(\e^{-\i s p}\eta)] + \sqrt{\gamma}\, \eta(-s) \sigma_+.
\end{align}
Therefore, if $\eta(x) = 0$ for all $x\in[-t,0]$, i.e.\ if the boson wavepacket is supported outside the space interval $[-t,0]$, then for all times $s\in[0,t]$ we get that 
\begin{equation}
     [H,B^\dag(\e^{-\i s p}\eta)] = [H_B,B^\dag(\e^{-\i s p}\eta)],
\end{equation}
whence $B^\dag(\eta,s)= B^\dag(\e^{-\i s p}\eta)$, for all $s\in[0,t]$, since they satisfy the same differential equation with the same initial condition. In particular, $B^\dag(\eta,t)= B^\dag(\e^{-\i t p}\eta)$.

In the case of a 2-point interaction we get instead
\begin{align}
    [H,B^\dag(\e^{-\i s p}\eta)] 
    &= [H_B,B^\dag(\e^{-\i s p}\eta)] 
    \nonumber\\
    & \quad + \sqrt{\frac{\gamma}{2}} \bigl(\,\eta(-s) + \eta(T-s)\,\bigr) \sigma_+,
\end{align}
and the free field evolution equals the Heisenberg evolution as far as $\eta(x) = 0$ for all $x\in[-t,0]\cup[T-t, T]$. 

\subsection{Choi state with nonlocal point interactions}\label{app:choi}
Let us evaluate the Choi state for a two-times process tensor on the Hamiltonian $H=H_0+H_{\mathrm{int}}$ with form factor given by Eq.~\eqref{eq:formfactor}, at times $(0,t_0,t_1)$ with $0<t_0<t_1<T$, starting from the system and the environment prepared in the global pure state $\ket{0,\vac}$. In particular, we will show that the Choi state is exactly the same that would be obtained with the form factor given by $g(x)=\sqrt{\gamma}\delta(x)$, which corresponds to a Markovian model, and therefore no deviation from Markovianity is detected if the process lasts less than $T$. The computation can be readily generalised for a general $k$-times process, again provided that the final time satisfies $t_{k-1}<T$.

Following Ref.~\cite{modi}, it suffices to consider two couples $S_0,S'_0$ and $S_1,S_1'$ of qubits each initially in a maximally entangled state $\ket{\Phi^+}_0,\ket{\Phi^+}_1$, defined by
\begin{equation}
\ket{\Phi^+}_s=\frac{1}{\sqrt{2}}\left(\ket{00}_s+\ket{11}_s\right),\;s=0,1.
\end{equation}
First of all, we must compute the state
\begin{equation}\label{eq:choi0}
	\ket{\Upsilon^{\mathrm{SE}}_{t_1+t_0,t_0,0}}=U^{(1)}_{t_1}U^{(0)}_{t_0}\ket{\Phi^+}_1\otimes\ket{\Phi^+}_0\otimes\ket{\vac},
\end{equation}
where $U^{(0)}_{t_0}$ is the evolution, for a time $t_0$, of the subsystem composed by the qubit $S'_0$ and the environment (with $S_0$ and the second couple of qubits $S_1,S_1'$ being uncoupled in the process) and $U^{(1)}_{t_1}$ is the evolution, for a time $t_1$, of the subsystem composed by the qubit $S'_1$ and the environment (with $S_1$ and the first couple of qubits $S_0,S_0'$ being uncoupled in the process). The partial trace of the pure state corresponding to the vector~\eqref{eq:choi0} with respect to the environment is the Choi state that we need to compute,
\begin{equation}\label{eq:choi}
	\Upsilon_{t_1+t_0,t_0,0}=\mathrm{tr}_{B}\left(\Ket{\Upsilon^{\mathrm{SE}}_{t_1+t_0,t_0,0}}\!\Bra{\Upsilon^{\mathrm{SE}}_{t_1+t_0,t_0,0}}\right).
\end{equation}
whose components may be arranged in a $16\times16$ matrix. Clearly, to compute~\eqref{eq:choi0}, we must evaluate the following four quantities:
\begin{eqnarray}
U^{(1)}_{t_1}U^{(0)}_{t_0}\ket{00}_1\otimes\ket{00}_0\otimes\ket{\vac};\\
U^{(1)}_{t_1}U^{(0)}_{t_0}\ket{00}_1\otimes\ket{11}_0\otimes\ket{\vac};\\
U^{(1)}_{t_1}U^{(0)}_{t_0}\ket{11}_1\otimes\ket{00}_0\otimes\ket{\vac};\\
U^{(1)}_{t_1}U^{(0)}_{t_0}\ket{11}_1\otimes\ket{11}_0\otimes\ket{\vac}.
\end{eqnarray}
We will report here the calculation of the first term; the remaining ones will follow similarly. First of all, in the time interval $[0,t_0]$ the qubit $S'_0$, in the state $\ket{0}_0$, interacts with the environment in the state $\ket{\vac}$, while all other emitters do not evolve. Recalling the discussion in the first subsection of this Appendix, we have
\begin{equation}
	U^{(0)}_{t_0}\ket{00,\vac}_0=a(t_0)\ket{00,\vac}_0+B^\dag(\xi_{t_0})\ket{01,\vac}_0
\end{equation}
with $\xi_{t_0}\equiv\xi(t_0,\cdot)$ as in Eq.~\eqref{eq:boson}, and therefore the global state of the system after a time $t_0$ will be given by
\begin{widetext}
\begin{equation}
	U^{(0)}_{t_0}\ket{00}_1\otimes\ket{00}_0\otimes\ket{\vac}=a(t_0)\ket{00}_1\otimes\ket{00}_0\otimes\ket{\vac}+\ket{00}_1\otimes\ket{01}_0\otimes B^\dag(\xi_{t_0})\ket{\vac},
\end{equation}
and thus
\begin{equation}
U^{(1)}_{t_1}U^{(0)}_{t_0}\ket{00}_1\otimes\ket{00}_0\otimes\ket{\vac}=a(t_0)\ket{00}_0\otimes U^{(1)}_{t_1}\ket{00,\vac}_1+\ket{01}_0\otimes U^{(1)}_{t_1}B^\dag(\xi_{t_0})\ket{00,\vac}_1.
\end{equation}
We must finally compute the two states $U^{(1)}_{t_1}\ket{00,\vac}_1$ and $U^{(1)}_{t_1}B^\dag(\xi_{t_0})\ket{00,\vac}_1$. The first one is computed exactly as in the previous step:
\begin{equation}
U^{(1)}_{t_1}\ket{00,\vac}_1=a(t_1)\ket{00,\vac}_1+B^\dag(\xi_{t_1})\ket{01,\vac}_1;
\end{equation}
In order to compute the second state, here we will finally use Eq.~\eqref{eq:commute3}, which holds as a direct consequence of Eq.~\eqref{eq:commute2} provided that $t_0+t_1<T$, as in our case.  We have
\begin{eqnarray}
U^{(1)}_{t_1}B^\dag(\xi_{t_0})\ket{00,\vac}_1=B^\dag(e^{-\i pt_1}\xi_{t_0})U^{(1)}_{t_1}\ket{00,\vac}_1&=&B^\dag(e^{-\i pt_1}\xi_{t_0})\left(a(t_1)\ket{00,\vac}_1+B^\dag(\xi_{t_2})\ket{01,\vac}_1\right)\nonumber\\
&=&a(t_1)B^\dag(e^{-\i pt_1}\xi_{t_0})\ket{00,\vac}_1+B^\dag(e^{-\i pt_1}\xi_{t_0})B^\dag(\xi_{t_1})\ket{01,\vac}_1\nonumber
\end{eqnarray}
We finally obtain
\pagebreak
\begin{eqnarray}\label{eq:calcolone}
	U^{(1)}_{t_1}U^{(0)}_{t_0}\ket{00}_1\otimes\ket{00}_0\otimes\ket{\vac}&=&\ket{00}_1\otimes\ket{00}_0\otimes a(t_0)a(t_1)\ket{\vac}\nonumber\\
	&+&\ket{01}_1\otimes\ket{00}_0\otimes a(t_0)B^\dag(\xi_{t_1})\ket{\vac} \nonumber\\
	&+&\ket{00}_1\otimes\ket{01}_0\otimes B^\dag(e^{-ipt_1}\xi_{t_0})a(t_1)\ket{\vac}+\nonumber\\
	&+&\ket{01}_1\otimes\ket{01}_0\otimes B^\dag(\xi_{t_1})B^\dag(e^{-\i pt_1}\xi_{t_0})\ket{\vac}.
\end{eqnarray}
Notice that, in the two-photon state in Eq.~\eqref{eq:calcolone}, the two creation operators \textit{commute}, since $\xi_{t_1}$ is supported in $[0,t_1]\cup [T,T+t_1]$ while $e^{-\i tp}\xi_{t_0}$ is supported in $[t_1,t_1+t_0]\cup[t_1+T,t_1+T+t_0]$, the supports being disjoint. 

The remaining three pieces can be computed similarly:
\begin{eqnarray}\label{eq:calcolone2}
U^{(1)}_{t_1}U^{(0)}_{t_0}\ket{00}_1\otimes\ket{11}_0\otimes\ket{\vac}&=&\ket{01}_1\otimes\ket{11}_0\otimes B^\dag(\xi_{t_1})a_0(t_0)\nonumber\\
&+&\ket{00}_1\otimes\ket{11}_0\otimes a(t_1)a_0(t_0)\ket{\vac};\\\label{eq:calcolone3}
U^{(1)}_{t_1}U^{(0)}_{t_0}\ket{11}_1\otimes\ket{00}_0\otimes\ket{\vac}&=&\ket{11}_1\otimes\ket{01}_0\otimes a_0(t_1)B^\dag(e^{-\i t_1p}\xi_{t_0})\ket{\vac}\nonumber\\
&+&\ket{11}_1\otimes\ket{00}_0\otimes a_0(t_1)a(t_0)\ket{\vac};\\\label{eq:calcolone4}
U^{(1)}_{t_1}U^{(0)}_{t_0}\ket{11}_1\otimes\ket{11}_0\otimes\ket{\vac}&=&\ket{11}_1\otimes\ket{11}_0\otimes a_0(t_1)a_0(t_0)\ket{\vac}.
\end{eqnarray}
\end{widetext}
The pattern is as follows. Only the qubit $S'_0$ and $S'_1$ may flip, and only the transition $\ket{0}\rightarrow\ket{1}$ is allowed; for each possible combination,
\begin{itemize}
	\item $\ket{1}\rightarrow\ket{1}$ will yield a phase factor $a_0(t_j)=e^{-\i \varepsilon_0t_j}$, $j=0,1$;
	\item $\ket{0}\rightarrow\ket{0}$ will yield an exponential factor $a(t_j)=e^{-(\i \varepsilon_0+\frac{\gamma}{2})t_j}$, $j=0,1$;
	\item $\ket{0}\rightarrow\ket{1}$ will yield a boson, i.e. an operator $B^\dag(\xi_{t_1})$ if the transition happens during the second step (a boson propagates starting from $t_0$ up to $t_0+t_1$), and $B^\dag(e^{-\i t_1p}\xi_{t_0})$ if it happens in the first step (a boson propagates from time $0$ to $t_0$, and then is rigidly translated for an additional time $t_1$).
\end{itemize} 
The sum of the terms in Eqs.~\eqref{eq:calcolone}--\eqref{eq:calcolone4} finally gives the vector in~\eqref{eq:choi0}. Now all matrix elements of the Choi state~\eqref{eq:choi} can be computed: since each of the addends corresponds to a distinct qubit state, we must simply evaluate the norms of all boson components of the states in Eqs.~\eqref{eq:calcolone}--\eqref{eq:calcolone4}. But recalling that, as long as $t<T$,
\begin{itemize}
	\item $a(t)$ has exactly the same expression that would be obtained in the reference model with $g(x)=\sqrt{\gamma}\,\delta(x)$, i.e. $|a(t)|^2=e^{-\gamma t}$;
	\item the boson wavefunction $\xi(t,x)$ has exactly the same norm, by Eq.~\eqref{eq:bosonnorm}, of the boson wavefunction $\varphi_{t}(x)$ that would be obtained in the reference model, i.e. $\|\xi(t,\cdot)\|^2=1-e^{-\gamma t}$, and similarly the scalar products between the wavefunctions $e^{-\i t_1p}\xi_{t_0}$ and $\xi_{t_1}$ are equally zero,
\end{itemize}
we conclude that, as long as we consider a process tensor lasting less than $T$, the Choi states for the reference model and the one with a nonlocal double interaction are exactly identical. More generally, this holds when taking into account any form factor as in Eq.~\eqref{eq:generic}.

Therefore, to prove our claim, it suffices to show explicitly that the reference model is Markovian in the sense of Eq.~\eqref{eq:markovchoi}. By a direct computation one has the following expressions for the Choi states of the model: the single-time Choi state $\Upsilon_{t_0,0}$ is a 4x4 matrix in the form
\begin{equation}
	\Upsilon_{t_0,0}=\begin{pmatrix}
	\e^{-\gamma t_0}&0&0&\e^{-\gamma t_0}\\
	0&1-\e^{-\gamma t_0}&0&0\\
	0&0&0&0\\
	\e^{-\gamma t_0}&0&0&1
	\end{pmatrix}
\end{equation}
while the two-times Choi state $\Upsilon_{t_1+t_0,t_0,0}$ is a 16x16 matrix that can be written, in a block diagonal structure, as
\begin{equation}
\Upsilon_{t_1+t_0,t_0,0}=\begin{pmatrix}
\Upsilon^{(00,00)}&0&0&\Upsilon^{(00,11)}\\
0&\Upsilon^{(01,01)}&0&0&\\
0&0&0&0&\\
\Upsilon^{(11,00)}&0&0&\Upsilon^{(11,11)}
\end{pmatrix},
\end{equation}
where
\begin{widetext}
\begin{eqnarray}
\Upsilon^{(00,00)}&=&\left(
\begin{array}{cccc}
e^{-\gamma (t_0 + t_1)} & 0 & 0 & e^{-\gamma \left(t_1+\frac{t_0}{2}\right)} \\
0 & e^{-\gamma t_1 } \left(1-e^{-\gamma t_0 }\right) & 0 & 0 \\
0 & 0 & 0 & 0 \\
e^{-\gamma \left(t_1+\frac{t_0}{2}\right)} & 0 & 0 & e^{-\gamma t_1 } \\
\end{array}
\right);\\
\Upsilon^{(00,11)}&=&\left(
\begin{array}{cccc}
e^{-\gamma \left(t_0 +\frac{ t_1 }{2}\right)} & 0 & 0 & e^{-\frac{1}{2}\gamma(t_0+t_1)} \\
0 & e^{-\frac{1}{2} \gamma t_1} \left(1-e^{-\gamma t_0 }\right) & 0 & 0 \\
0 & 0 & 0 & 0 \\
e^{-\frac{1}{2}\gamma(t_0+t_1)} & 0 & 0 & e^{-\frac{1}{2} \gamma t_1} \\
\end{array}
\right);\\
\Upsilon^{(01,01)}&=&\left(
\begin{array}{cccc}
e^{-\gamma t_0 } \left(1-e^{-\gamma t_1 }\right) & 0 & 0 & e^{-\frac{1}{2} \gamma t_0} \left(1-e^{-\gamma t_1 }\right) \\
0 & \left(1-e^{-\gamma t_0 }\right) \left(1-e^{-\gamma t_1 }\right) & 0 & 0 \\
0 & 0 & 0 & 0 \\
e^{-\frac{1}{2} \gamma t_0} \left(1-e^{-\gamma t_1 }\right) & 0 & 0 & 1-e^{-\gamma t_1 } \\
\end{array}
\right);\\
\Upsilon^{(11,00)}&=&\left(
\begin{array}{cccc}
e^{-\gamma \left(t_0 +\frac{ t_1 }{2}\right)} & 0 & 0 & e^{-\frac{1}{2}\gamma(t_0+t_1)} \\
0 & e^{-\frac{1}{2} \gamma t_1} \left(1-e^{-\gamma t_0 }\right) & 0 & 0 \\
0 & 0 & 0 & 0 \\
e^{-\frac{1}{2}\gamma(t_0+t_1)} & 0 & 0 & e^{-\frac{1}{2} \gamma t_1} \\
\end{array}
\right);\\
\Upsilon^{(11,11)}&=&\left(
\begin{array}{cccc}
e^{-\gamma t_0 } & 0 & 0 & e^{-\frac{1}{2} \gamma t_0} \\
0 & 1-e^{-\gamma t_0 } & 0 & 0 \\
0 & 0 & 0 & 0 \\
e^{-\frac{1}{2} \gamma t_0} & 0 & 0 & 1 \\
\end{array}
\right).
\end{eqnarray}
\end{widetext}
With these expressions, a direct computation shows that the Choi states of the reference model satisfy the Markov property:
\begin{equation}
\Upsilon_{t_1+t_0,t_0,0}=\Upsilon_{t_1+t_0,t_0}\otimes\Upsilon_{t_0,0}.
\end{equation}
This calculation is immediately generalisable to a process tensor with a larger number of steps.

\bibliography{refs}

\end{document}